# Sample Designs and Estimators for Multimode Surveys With Face-to-Face Data Collection

J. Michael Brick and Jill M. DeMatteis[1]

## Abstract

Survey researchers are increasingly turning to multimode data collection to deal with declines in survey response rates and increasing costs. An efficient approach offers the less costly modes (e.g., web) followed with a more expensive mode for a subsample of the units (e.g., households) within each primary sampling unit (PSU). We present two alternatives to this traditional design. One alternative subsamples PSUs rather than units to constrain costs. The second is a hybrid design that includes a clustered (two-stage) sample and an independent, unclustered sample. Using a simulation, we demonstrate the hybrid design has considerable advantages.

---

[1] J. Michael Brick, Statistics and Data Science, Westat, 1600 Research Blvd., Rockville, MD 20850 USA; Jill M. DeMatteis, Statistics and Data Science, Westat, 1600 Research Blvd., Rockville, MD 20850 USA

## Section 1. Introduction

Many surveys use web and mail data collection modes due to their relatively low costs, but the overall response rates may be lower than desired, and the estimates may be subject to considerable nonresponse bias (Dillman 2017; Brick et al. 2021). Introducing face-to-face (ftf) interviewing as a follow-up mode can substantially increase response rates, although at a higher cost. The increased response rates achieved through ftf follow-up typically also result in reduced nonresponse bias, due to the introduction of the ftf mode resulting in increased response propensities for subgroups that are less likely to respond to the web or mail modes. A sampling approach to make multimode data collection with ftf interviewing more cost-efficient is two-phase sampling. Hansen and Hurwitz (1946) introduced two-phase sampling as a method of dealing with nonresponse based on the two-phase sampling theory of Neyman (1938). Hansen and Hurwitz applied their method with a first phase sample of retail establishments that were sent a mail questionnaire, with non-respondents subsampled for ftf. Since their ftf efforts resulted in virtually 100% response after weighting for subsampling, the estimates from the survey were unbiased.

A well-known application of this sampling strategy is the American Community Survey (ACS) where a sample of households is selected within tabulation areas and the households are requested to respond to a web or mail questionnaire; a subsample of nonrespondents within the area are followed up ftf (U.S. Census Bureau 2014; U.S. Census Bureau 2019). Like other surveys today, the ACS does not achieve full response so the survey weights must be further adjusted for nonresponse and the estimates are still subject to potential nonresponse bias.

This article examines two new sample designs for multimode household surveys that uses ftf interviewing to increase response rates. Surveys with this type of multimode design may be more common in the U.S. and Canada. The standard two-phase sampling approach selects a sample of primary sampling units (PSUs) and households within the PSU for the low-cost mode(s) and then subsamples within each PSU for ftf follow-up. The two new approaches alter the designs for the low cost (web/mail) to improve the efficiency of the estimates while retaining the basic ftf design. Designs that use more expensive modes before the lower cost modes are not considered (e.g., Bayart and Bonnel 2015). The first new sample design we explore is a variant where a large sample of PSUs is sampled and households in each PSU are recruited by low-cost modes in the first phase, but the ftf follow-up is only done in a subsample of PSUs. This approach reduces the clustering effect by spreading the respondents from the first phase sample over a larger number of PSUs. We also consider a second approach that selects two independent samples – one an unclustered sample of households that is recruited by low-cost modes and a second clustered sample where all sampled households are recruited sequentially using all modes. The estimates are created by compositing the data from the two samples. We do not discuss nonsampling errors that may arise when more than one mode data collection is used (e.g., Goodman et al. 2022).

In Section 2, we describe the two new sample designs in more detail. We present estimators and describe their properties under different nonresponse models in Section 3. Section 4 describes and gives the results of a simulation study we conducted. We conclude in Section 5 with some discussion of the implications of the design options and estimators, recommendations, and areas for future research.

## Section 2. Sampling Designs

All the applications of two-phase household sampling for nonresponse where ftf interviewing is used that we have identified begin with a first phase sample of PSUs (clustered geographically) and then subsample nonrespondents for the second phase from each of the first phase PSUs. In many cases, the subsampling for nonresponse is an adaptive or responsive design feature rather than an initial design approach (Groves and Heeringa, 2006; Heeringa et al. 2004; Wagner et al. 2014). The first phase is a sample of PSUs with a sample of households from an address-based sampling (ABS) frame. All sampled households are subject to the initial data collection protocol that may involve web and/or mail as a low-cost mode.

***Two-phase unit subsampling.*** The nonrespondents to the first-phase sample are subsampled within each PSU and ftf interviewing is the mode for the second phase. The ACS uses this design. We refer to this standard approach as *two-phase unit subsampling*. Särndal and Swensson (1987) extended the theory of Hansen and Hurwitz to designs where the first phase sample was not a simple or stratified random sample.

Two statistical issues arise in two-phase unit subsampling design. One concern is the increase in variance of estimates due to subsampling (Kish 1991) since the weights account for the subsampling. The second issue is the increase in variance due to clustering, assuming a positive intraclass correlation for the characteristics of interest, because all responses are clustered within the sampled PSUs. In addition, a related cost and operational issue is limiting the number of sampled PSUs to make the ftf data collection within the PSUs effective.

***Two-phase PSU subsampling.*** An alternative two-phase design is to select a large number of PSUs and households in the first phase and then select a subsample of the PSUs and all of the households within those PSUs for the second phase. All households sampled in the first phase are recruited by web, but only the nonrespondents in the subsample of PSUs are followed up in the second phase by ftf. We refer to this approach as *two-phase PSU subsampling*. We have not identified any surveys using this approach. This approach has benefits and concerns similar to those with two-phase unit subsampling, but has the advantage of reducing the clustering effect by spreading the respondents from the first phase sample over a larger number of PSUs.

***Hybrid sampling.*** A different design with fewer restrictions than two-phase PSU subsampling selects two independent samples and then composites the estimates from the two samples to produce final estimates. The first sample is an unclustered sample of households from the ABS frame and those households are recruited only by low-cost modes. The second sample uses a two-stage design to reduce data collection costs, with a sample of PSUs and households within the PSUs. The data are collected from households in this second sample by sequentially using web and then ftf modes. This design builds on both dual frame (Lohr 2011) and two-phase methods. Like dual frame methods it selects two samples, although from the same frame in this case, and then combines the data from the two samples. Like two-phase sampling, only a subsample of the full sample of households is subject to the full data collection protocol. We refer to this as hybrid two-phase sampling or more concisely as *hybrid sampling*. When discussing the two samples, we refer to them as the *unclustered* and *clustered* samples, respectively, for ease of discussion.

***Example.*** Here, we present an illustration of each design and consider the effects on precision; later, for each of these design alternatives, we will examine the bias of various estimators. As a simple example of the three approaches, suppose the goal is to complete 10,000 household interviews with 70 percent done by web and 30 percent by ftf. For illustration, we ignore nonresponse weighting adjustments and details about costs that are discussed later.

The three approaches for this simple example are illustrated in Exhibit 2.1. The two-phase unit subsampling approach selects a sample of 200 PSUs with probability proportionate to the number of households, and then an equal number of households is selected in each PSU. All sampled households are pushed to web and a subsample of web nonrespondents in each PSUs are followed by ftf. Assuming a 25% web response rate and 50% ftf response rate, then sampling 140 households per PSU yields 35 web completes. Subsampling roughly 30 of the approximately 105 nonrespondents per PSU results in 15 ftf completes for a total of 50 responses (35 by web and 15 by ftf). The design effect due to subsampling (differential weighting) is approximately 1.44 (Kish 1992) and reaches this maximum value when the outcome and the weights are uncorrelated. The design effect due to clustering is approximately $1 + \delta(\bar{m} - 1)$, where $\bar{m}$ is the average number of completed households per PSU and $\delta$ is the intraclass correlation. With 50 completes per PSU and $\delta = .02$, the clustering design effect is about 2.0. The overall design effect is the product of the weighting and clustering effects, or 2.9, and an effective sample size of about 3,500 (10,000/2.9).

**Exhibit 2.1. Illustration of three follow-up approaches**

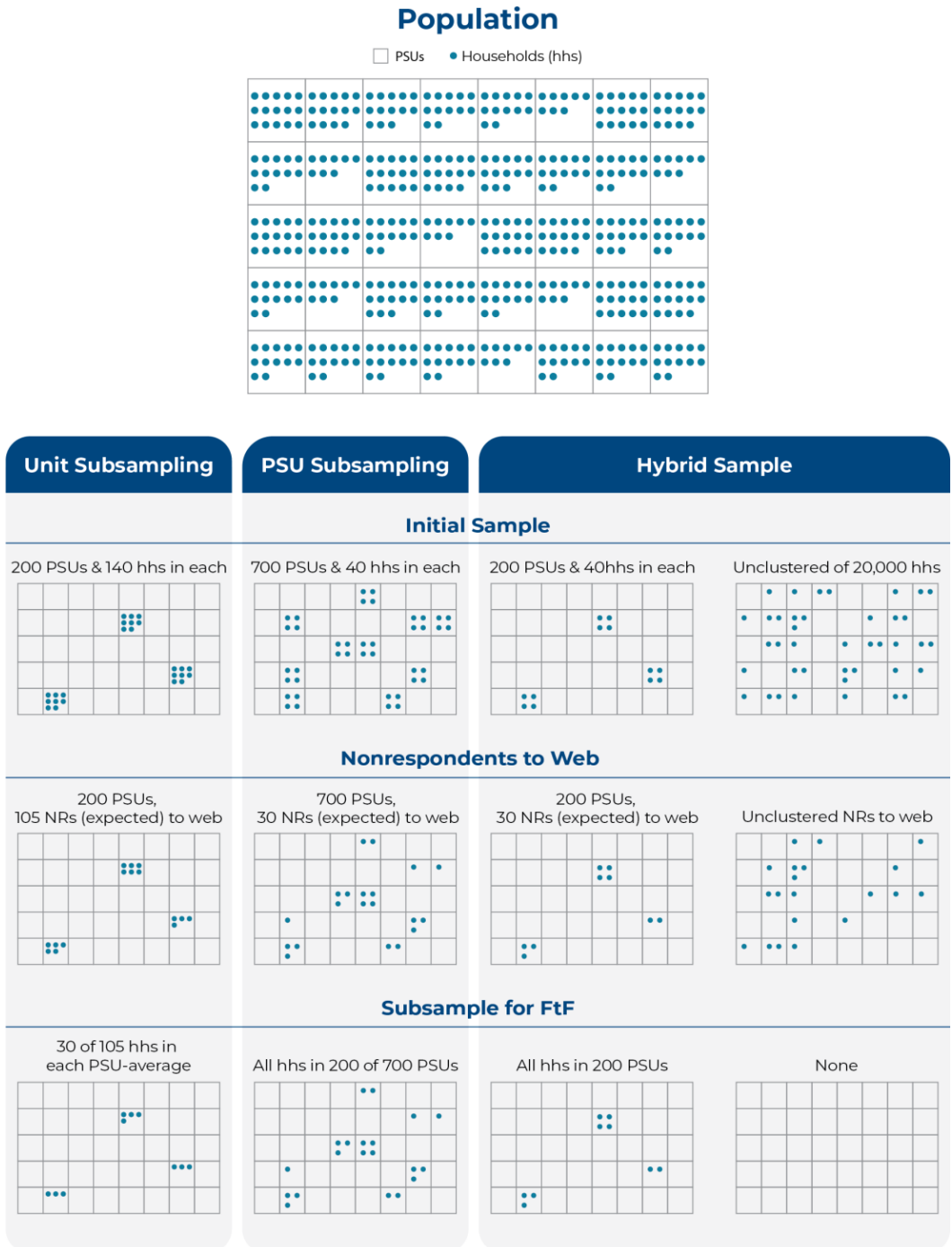

In the two-phase PSU subsampling approach, a sample of 700 PSUs is selected with probability proportional to the number of households, and 200 PSUs are subsampled with equal probability. The number of PSUs for the subsample was chosen to equal the number in the unit subsampling design, and 700 total PSUs allows taking all nonrespondents within the 200 subsampled PSUs without further subsampling. In each PSU an equal probability sample of 40 households is selected. All the sampled households are sent to web, resulting in approximately 7,000 web completes (700*40*.25). All of the web nonrespondents in the 200 PSU subsample are followed up ftf, yielding 3,000 ftf completes. The design effect due to weighting is still 1.44. Due to the unequal number of completes per cluster, we use the approximate design effect suggested by Holt (1980), with $m' = \sum m_i^2 / \sum m_i$ instead of $\bar{m}$ where $m_i$ is the number of completes in PSU $i$. In this design $m_i = 10$ in 500 PSUs and is $m_i = 25$ in 200 PSUs, so $m' = 17.5$ and the design effect due to clustering is $1 + \delta(m' - 1) = 1.33$ when $\delta = .02$. The overall design effect is 1.9 and the effective sample size is just over 5,200, a substantial increase in precision over the traditional two-phase approach.

With the hybrid approach a sample of 200 PSUs (equal to the number of PSUs where ftf is done in the other designs) is selected and 40 households per PSU in the clustered sample, yielding 2,000 web completes and 3,000 ftf completes. The unclustered sample is 20,000 to yield 5,000 web completes. The design effect due to clustering in the clustered sample is 1.48 when $\delta = .02$ and there is no differential weighting effect (assuming a uniform nonresponse adjustment is applied to all respondents). In all, there are 7,000 web and 3,000 ftf completes. The two samples are composited using, for example, $\lambda = .7$ for the unclustered sample (since it 70% of the total) and $1 - \lambda = .3$ for the clustered sample, so the overall design effect is 1.14. (An optimal compositing factor, which also takes into account the clustering effect on precision for the clustered sample, could be determined and applied; however, for simplicity in this illustration, we chose compositing factors proportional to the number of completes from the particular mode. We examine the effects of alternative compositing factors later in the manuscript.) The effective sample size is 8,770, a substantial increase in precision over both two-phase approaches.

## Section 3. Estimation and Nonresponse Models

### Nonresponse Models

In the literature on models of survey nonresponse, two frameworks have emerged: a *deterministic* framework that partitions the population into two mutually exclusive, exhaustive groups, respondent and nonrespondents (as described on pp. 359-363 of Cochran 1977); and a *stochastic* framework in which each member of the population has a probability of responding to a particular survey (see, for example, Brick and Montaquila 2009).

Using the deterministic view of nonresponse under a given data collection protocol, the population contains a set of households who will respond to the web request, a set of households that will respond to the ftf request after not responding on the web, and a set of households that will not respond at all. With this model, the population total is

$$Y = N\left[\gamma_W \bar{Y}_W + \gamma_F \bar{Y}_F + (1 - \gamma_W - \gamma_F)\bar{Y}_N\right], \quad (3.1)$$

where $\gamma_W$ is proportion responding by web, $\gamma_F$ is proportion responding by ftf but not by web, and $\bar{Y}_W$, $\bar{Y}_F$, and $\bar{Y}_N$ are the corresponding population means of the characteristic of interest for the web, ftf, and nonresponding sets. The deterministic view with the fixed partitioning of the population by response mode under the data collection protocol is conceptualized from a post-data collection perspective, where repeated implementations of the same protocol could be used to determine these constants.

A stochastic model that aligns with this deterministic model assumes each unit in the population has a response propensity vector with the first element the probability of responding by web and the second element the probability of responding by ftf and not by web, say $\phi_k = (\phi_{k,W}, \phi_{k,F})'$, where $0 \leq \phi_{k,W} \leq 1; 0 \leq \phi_{k,F} \leq 1; 0 < \phi_{k,W} + \phi_{k,F} \leq 1$. Note that if we define $\phi_{k,F|W^c}$ to be the conditional probability unit $k$ responds by ftf given that they did not respond by web, then $\phi_{k,F} = (1 - \phi_{k,W})\phi_{k,F|W^c}$. In addition, taking expectations over this response distribution gives

$$E_R \sum_{k \in U} \phi_{k,W} \equiv \gamma_W N; E_R \sum_{k \in U} \phi_{k,W} y_k \equiv Y_W \text{ where } U \text{ denotes the population, } Y_W = \gamma_W N \bar{Y}_W,$$

and $E_R \sum_{k \in U} \phi_{k,F} \equiv \gamma_F N; E_R \sum_{K \in U} \phi_{k,F} y_k \equiv Y_F$ where $Y_F = \gamma_F N \bar{Y}_F$. With full response (i.e., $\phi_{k,F|W^c} = 1$, so that $\phi_{k,W} + \phi_{k,F} = 1 \forall k$), we have $Y = N[\gamma_W \bar{Y}_W + \gamma_F \bar{Y}_F]$.

**Estimators**

First, suppose we follow all web nonrespondents by ftf, i.e., a two-stage sample with no subsampling. The typical approach to estimation is to use the Horvitz-Thompson (HT) estimator with an adjustment for nonresponse. (Here, we consider a single overall adjustment, ignoring more complex nonresponse and calibration adjustments for simplicity.) This estimator of the total is

$$\hat{t}_1 = \sum_{k \in S} d_k \delta_k(W) \hat{R}^{-1} y_k + \sum_{k \in S} d_k \delta_k(F) \hat{R}^{-1} y_k$$

$$= \hat{N} \left[ \frac{\hat{\gamma}_W}{\hat{\gamma}_W + \hat{\gamma}_F} \bar{y}_W + \frac{\hat{\gamma}_F}{\hat{\gamma}_W + \hat{\gamma}_F} \bar{y}_F \right] \quad , \quad (3.2)$$

where $S$ denotes the sample, $d_k$ is the reciprocal of the probability of selection of household $k$ (accounting for both stages of selection), $\delta_k(W) = 1$ if household $k$ responds to the web survey and is 0 otherwise, $\delta_k(F) = 1$ if household $k$ responds to the ftf survey and is 0 otherwise, $\hat{N} = \sum_{k \in S} d_k$, and $\hat{R} = \hat{R}_W + (1 - \hat{R}_W)\hat{R}_F$ where $\hat{R}_W = \sum_{k \in S} d_k \delta_k(W) / \sum_{k \in S} d_k$ and $\hat{R}_F = \sum_{k \in S} d_k \delta_k(F) / \sum_{k \in S} d_k (1 - \delta_k(W))$ are the observed web and ftf response rates. Note that $\hat{R}_F$ is the conditional response rate given no response to web. Throughout, we estimate $\hat{\gamma}_W = \hat{R}_W$ and $\hat{\gamma}_F = (1 - \hat{R}_W)\hat{R}_F$ by the observed response rates where those depend upon the specific design.

The estimator $\hat{t}_1$ is unbiased if $\bar{Y}_N = \bar{Y}_F = \bar{Y}_W$ or if the ftf conditional response rate is 100%. This property is easily shown by taking expectations with respect to both the sampling and response distributions (Särndal and Swensson, 1987). We assume throughout that standard conditions for the appropriate full response estimator to be unbiased also hold.

In contrast to $\hat{t}_1$, which applies a constant adjustment $\hat{R}^{-1}$ to all respondents, we also consider an estimator commonly used in the two-phase sampling context, which adjusts only the ftf respondents, as follows:

$$\begin{aligned}\hat{t}_2 &= \sum_{k \in S} d_k \delta_k(W) y_k + \sum_{k \in S} d_k \delta_k(F) \hat{R}_F^{-1} y_k \\ &= \hat{N}\left[\hat{\gamma}_W \bar{y}_W + (1-\hat{\gamma}_W)\bar{y}_F\right]\end{aligned} \quad (3.3)$$

The estimator $\hat{t}_2$ is unbiased if $\bar{Y}_N = \bar{Y}_F$ or if the ftf conditional response rate is 100%.

Next, we consider the subsampling designs and extend these estimators to incorporate the subsampling. Let $\omega$ denote the conditional probability of selection into the subsample for ftf interviewing. With two-phase unit subsampling with the second phase subsampling rate of $\omega$ ($0 < \omega \leq 1$), the extensions of these estimators are

$$\begin{aligned}\hat{t}_1 &= \sum_{k \in S} d_k \delta_k(W) \hat{R}^{-1} y_k + \sum_{k \in S} d_k \omega^{-1} \delta_k(F) \hat{R}^{-1} y_k \\ &= \hat{N}\left[\frac{\hat{\gamma}_W}{\hat{\gamma}_W + \hat{\gamma}_F}\bar{y}_W + \frac{\hat{\gamma}_F}{\hat{\gamma}_W + \hat{\gamma}_F}\bar{y}_F\right]\end{aligned} \quad (3.4)$$

and

$$\begin{aligned}\hat{t}_2 &= \sum_{k \in S} d_k \delta_k(W) y_k + \sum_{k \in S} d_k \omega^{-1} \delta_k(F) \hat{R}_F^{-1} y_k \\ &= \hat{N}\left[\hat{\gamma}_W \bar{y}_W + (1-\hat{\gamma}_W)\bar{y}_F\right]\end{aligned} \quad (3.5)$$

In this expression the subsampling rate is accounted for in estimating $\gamma_W$ (Exhibit 3.1 shows how this and other estimators are written in terms of weights). Note that the estimators given in equations (3.2) and (3.3) are a special case of those in equations (3.4) and (3.5), respectively, where $\omega = 1$.

For the subsampling designs, both estimators are unbiased if the ftf conditional response rate is 100%; $\hat{t}_2$ is unbiased if $\bar{Y}_N = \bar{Y}_F$, while $\hat{t}_1$ requires the more stringent condition that $\bar{Y}_N = \bar{Y}_F = \bar{Y}_W$. This result applies to both two-phase unit and PSU subsampling. While Särndal and Swensson (1987) did not consider two-phase PSU subsampling, their proof of the unbiasedness applies since the subsampling they consider is not specific to a phase. Thus, the two-phase extension of the estimator $\hat{t}_2$ shown in equation (3.5) can be used with either two-phase unit subsampling or two-phase PSU subsampling.

In the PSU subsampling design, the weighting adjustment in $\hat{t}_2$, $\omega^{-1}$, is the ratio of the number of first phase PSUs to the number of PSUs in the second phase. An alternative subsampling adjustment is

$$\omega_s^{-1} = \frac{\sum_{k \in S} d_k (1 - \delta_k(W))}{\sum_{k \in \text{sub-PSU}} d_k (1 - \delta_k(W))}, \qquad (3.6)$$

where the numerator is the sum of the weights of all sampled cases that did not respond by web and the denominator is the sum of the weights of the web nonrespondents in the subsampled PSUs. This adjustment incorporates the size of the subsampled PSUs, where size is the number of subsampled households.

For hybrid sampling, we use dual frame notation to simplify the presentation. Let $S_A$ be an unclustered sample of households, and $S_B$ be a two-stage sample with households sampled within PSUs. The protocol for $S_A$ just uses web, while for $S_B$ web is followed by ftf for all web nonrespondents.

Since $S_A$ is a single stage web data collection, the Horvitz-Thompson estimator or (3.2) with only web respondents is

$$\hat{t}_A = \frac{\sum_{k \in S_A} d_k}{\sum_{k \in S_A} d_k \delta_k(W)} \sum_{k \in S_A} d_k \delta_k(W) y_k = \hat{N}_{S_A} \bar{y}_{WA}, \qquad (3.7)$$

where $\hat{N}_{S_A} = \sum_{k \in S_A} d_k$ and $\bar{y}_{WA}$ is the estimated mean for web respondents based on the unclustered sample $S_A$. This *unclustered* estimator is unbiased if $\bar{Y}_N = \bar{Y}_F = \bar{Y}_W$ as discussed for $\hat{t}_1$.

Rewriting $\hat{t}_1$ for a clustered sample $S_B$ gives

$$\hat{t}_{B,1} = \hat{N}_{S_B} \left[ \frac{\hat{\gamma}_W}{\hat{\gamma}_W + \hat{\gamma}_F} \bar{y}_{WB} + \frac{\hat{\gamma}_F}{\hat{\gamma}_W + \hat{\gamma}_F} \bar{y}_{FB} \right], \qquad (3.8)$$

where $\hat{N}_{S_B} = \sum_{k \in S_B} d_k$, and $\bar{y}_{WB}$ and $\bar{y}_{FB}$ are the estimated means for web respondents and ftf respondents, respectively, based on the clustered sample $S_B$.

Compositing the two estimators for the hybrid samples gives (assume $\hat{N} = \hat{N}_{S_A} = \hat{N}_{S_B}$)

$$\hat{t}_{df,1} = \lambda \hat{t}_A + (1-\lambda)\hat{t}_{B,1}$$
$$= \hat{N}\left[\lambda \bar{y}_{WA} + (1-\lambda)\frac{\hat{\gamma}_W}{\hat{\gamma}_W + \hat{\gamma}_F}\bar{y}_{WB} + (1-\lambda)\frac{\hat{\gamma}_F}{\hat{\gamma}_W + \hat{\gamma}_F}\bar{y}_{FB}\right]. \quad (3.9)$$

This estimator, like (3.2), does not place more weight on the ftf respondents and is unbiased when $\bar{Y}_N = \bar{Y}_F = \bar{Y}_W$ as discussed previously. The more stringent nonresponse model is required for unbiasedness because while $\hat{t}_{B,1}$ estimates the total for the population consisting of both the web and ftf domains, $\hat{t}_A$ estimates the total for only the web domain.

Another option is to composite two estimators but to use the two-phase version of (3.5) for the clustered sample. First, we composite the web respondents from $S_A$ and $S_B$ and then adjust the weight of ftf respondents to account for the remaining nonresponse. This composite estimator (where only the web samples are composited) is

$$\hat{t}_{df,2} = \hat{N}\hat{\gamma}_W\left[\kappa \bar{y}_{WA} + (1-\kappa)\bar{y}_{WB}\right] + \hat{N}(1-\hat{\gamma}_W)\bar{y}_{FB}. \quad (3.10)$$

This estimator is unbiased when $\bar{Y}_N = \bar{Y}_F$ or when the ftf conditional response rate is 100% using the same arguments as for (3.5). This estimator is efficient since it includes both the clustered and unclustered sample observations. The estimate of $\hat{\gamma}_W$ is the observed proportion of web respondents based on both samples.

**Exhibit 3.1.    Estimators, weights and nonresponse models for two-phase and hybrid sampling**

| Estimator | Design | Respondent weight | | Nonresponse model |
|---|---|---|---|---|
| $\hat{t}_1$ | Two stage | $d_k \hat{R}^{-1}$ | Web respondents in $S_B$ | $\bar{Y}_N = \bar{Y}_F = \bar{Y}_W$ |
| | | $d_k \omega^{-1} \hat{R}^{-1}$ | Ftf respondents in $S_B$ | |
| $\hat{t}_2$ | Two phase | $d_k$ | Web respondents in $S_B$ | $\bar{Y}_N = \bar{Y}_F$ |
| | | $d_k \omega^{-1} \hat{R}_F^{-1}$ | Ftf respondents in $S_B$ | |
| $\hat{t}_A$ | One stage, unclustered | $d_k \hat{R}_W^{-1}$ | All (web) respondents in $S_A$ | $\bar{Y}_N = \bar{Y}_F = \bar{Y}_W$ |
| $\hat{t}_{df,1}$ | Hybrid | $\lambda d_k \hat{R}_W^{-1}$ | All (web) respondents in $S_A$ | $\bar{Y}_N = \bar{Y}_F = \bar{Y}_W$ |
| | | $(1-\lambda) d_k \hat{R}^{-1}$ | All respondents in $S_B$ | |
| $\hat{t}_{df,2}$ | Hybrid | $\kappa d_k$ | All (web) respondents in $S_A$ | $\bar{Y}_N = \bar{Y}_F$ |
| | | $(1-\kappa) d_k$ | Web respondents in $S_B$ | |
| | | $d_k \hat{R}_F^{-1}$ | Ftf respondents in $S_B$ | |

If the assumed nonresponse model is $\bar{Y}_N = \bar{Y}_F$ or with full ftf response, then $\hat{t}_2$ is an unbiased estimator when the design is either two-phase unit subsampling or two-phase PSU subsampling and $\hat{t}_{df,2}$ is an unbiased estimator when the design is hybrid sampling. More efficient estimators ($\hat{t}_1$ and $\hat{t}_{df,1}$) are available but require the more stringent assumption $\bar{Y}_N = \bar{Y}_F = \bar{Y}_W$.

The only remaining parameters that need to be specified for the hybrid estimators are the compositing factors. The usual approach is a composite factor $\lambda$ ($0 \leq \lambda \leq 1$) equal to the effective relative sample size and this is often a reasonable approximation. For $\hat{t}_{df,1}$, $\lambda$ might be set to be the ratio of the effective number of respondents in $S_A$ divided by the sum of that and the effective number of respondents in $S_B$, where the design effect for $S_B$ is estimated as described in the earlier example. A similar approach could be used for $\kappa$, but since only the web respondents are being composited, the effective sample sizes are only those for the web respondents. We explore different compositing factors in the simulation and find that reasonable choices have little effect on the variances of the estimates.

## Variance Estimation

Variance estimation methods for most of the estimators are well-known or require only minor adjustments to handle the approaches proposed here for nonresponse follow-up (i.e., two-stage, unclustered estimators and hybrid sampling). For example, hybrid sampling is covered by dual-frame estimation theory (Lohr 2011). Variance estimation for two-phase sampling with complex sampling schemes has been the subject of several recent theoretical developments including Hidiroglou (2001), Hidiroglou, Rao and Haziza (2009) and Beaumont, Béliveau and Haziza (2015).

Variance estimation for two-phase PSU subsampling requires some extensions of Beaumont, Béliveau and Haziza (2015) to ensure the variances are appropriately estimated. The only design we considered here involves sampling a large number of PSUs in the first phase and then taking all the nonrespondents in the second phase in a subsample of the PSUs. The subsampling of PSUs for the second phase does not depend on the outcomes of the first phase sample (we also assume that every sampled first phase PSU has some nonrespondents). Since this design selects all the nonrespondents in the subsampled PSU it satisfies the invariance and independence conditions (see Särndal, Swensson and Wretman 1992. p. 134-135) for two-stage sampling. Provided the first stage sampling fraction is negligible, the standard with replacement variance estimation can be applied.

We used Taylor series linearization for our simulation but Beaumont, Béliveau and Haziza (2015) describe how replication methods apply equally well in this situation. The sample design for two-phase PSU subsampling typically selects a stratified sample of PSUs. (Our simulation used an unstratified sample, but the generalization to a stratified sample is straightforward.) In each stratum, the number of PSUs selected is determined so that the units treated as variance strata (or clusters) are balanced with respect to the subsampling for follow-up. This combining of PSUs gives an unbiased estimate of the variance (Wilson, Brick and Sitter 2006). For example, with a 50% subsample of PSUs for the second phase, one approach is to sample 4 PSUs in each stratum and pair these to form 2 variance units so that each variance unit has one PSU subsampled for follow-up and one with no follow-up. With a 33% subsample of PSUs, 6 PSUs are sampled per stratum in the first phase and 2 variance units are formed and each contains 2 PSUs not subsampled for follow-up and 1 subsampled for follow-up.

## Section 4.  Simulation Study

To examine the three sample designs and estimators, we conducted a simulation study. In this section, we begin by describing the approach we used to generate the populations for this simulation. Next, we lay out our simulation design, including a description of the measures we used to evaluate and compare the designs and estimators. Finally, we present the simulation results, beginning with those for the hybrid design scenarios, followed by a comparison of the subsampling designs to the hybrid design.

## Generating the Population

For this simulation, we used the 2015-2019 ACS Public Use Microdata Sample (PUMS) data for the 50 States and D.C., treating all ACS respondents residing in households as our population. ACS data collection begins with a mailed invitation to complete the survey on the *web*. Nonrespondents are followed up first by mailing a paper questionnaire (*mail*), then by computer- assisted telephone interviewing (CATI) or computer-assisted personal interviewing (CAPI) for a subsample of nonrespondents (*ftf*). (Although CATI was dropped as a follow-up mode for ACS after September 2017, that is not relevant to our simulation, as all CATI and CAPI respondents are treated as ftf respondents for our simulation purposes.) On the ACS PUMS files, each household is associated with a geographic cluster called a Public Use Microdata Area (PUMA). We used PUMAs as primary sampling units (PSUs) for the clustered samples. In selecting clustered samples, we sampled PSUs with probabilities proportional to the number of households in the PSU, and sampled households with equal probabilities within PSUs, where the conditional probabilities of selection were proportional to the reciprocal of the PSU selection probabilities. For each unclustered sample, we sampled households using a simple random sample without replacement.

In the previous section, we discussed the nonresponse model under which the estimator is unbiased for each estimator. To examine each estimator's performance under various response models, we created four different pseudopopulations (A, B, C, and D) by varying the approach used to identify respondents and nonrespondents. For pseudopopulations A, B, and C, we considered all ACS web respondents to be our web-push respondents. For pseudopopulation A, we designated all ACS mail respondents to be our ftf respondents, and all ACS ftf respondents to be our nonrespondents. For pseudopopulation B, among the ACS mail and ftf respondents, we randomly identified half to be our ftf respondents and the other half to be our nonrespondents. Pseudopopulation C has a 100% response with the ftf follow-up, where all ACS mail and ftf respondents were our ftf respondents. Our motivation for constructing pseudopopulation D was to create a population with a lower response rate to the initial mode of contact; thus, we defined all ACS mail respondents as our web respondents, and as we did in creating pseudopopulation B, we randomly identified half of the remainder to be our ftf respondents and the other half to be our nonrespondents.

We included several demographic and socioeconomic variables at the household level in this simulation. Figure 4.1 gives brief descriptions of each of the variables and displays the population means of each of the variables by ACS mode of completion, where the full ACS PUMS dataset is the population. For pseudopopulation A, we expect all estimators to be biased to some extent since the nonrespondents differ from both groups of respondents. The ftf follow-up, when properly accounted for in the estimator, would be expected to reduce bias in estimates for variables v1, v3, v8, and v11 based on Figure 4.1. For pseudopopulations B and D, the response model $\bar{Y}_N = \bar{Y}_F$ holds by design, so estimators based on that response model should be unbiased. For pseudopopulation C, there is no nonresponse after the ftf follow-up so estimators that appropriately weight the follow-up should be unbiased.

**Simulation Design**

The simulations involved selecting samples from the various pseudopopulations (depicted in Exhibit 4.1) and, for each variable, computing estimated totals using each of the estimators presented in Section 3. The simulation scenarios are listed in Exhibit 4.2. The scenarios are labeled according to the combination of parameters; e.g., scenario B2U is the scenario that uses pseudopopulation B, selects a clustered sample only, and uses two-phase unit subsampling for ftf nonresponse follow-up. For scenarios A1A, B1A, C1A, and D1A, each of which examines the hybrid design with different pseudopopulations, we selected clustered and unclustered samples of size 2,500 each, with 50 PSUs for the clustered sample. To allow for a direct comparison of the hybrid design approach (specifically, scenario B1A) to the unit subsampling and PSU subsampling approaches, we included scenarios B2P and B2U.

**Figure 4.1.** Population means for variable from American Community Survey 2015-2019 PUMS, by mode. Of all responses, web is 49.3%, mail is 29.1%, and ftf is 21.7%.

Variable definitions: v1-high school graduate; v2-some college; v3-Bachelor's or higher; v4-1-person household; v5-2-person household; v6-3 or more person household; v7-Hispanic reference person; v8-Black, nonHispanic reference person; v9-renter; v10-person 60 years or older; v11*-household income/110000; v12-percent urban.

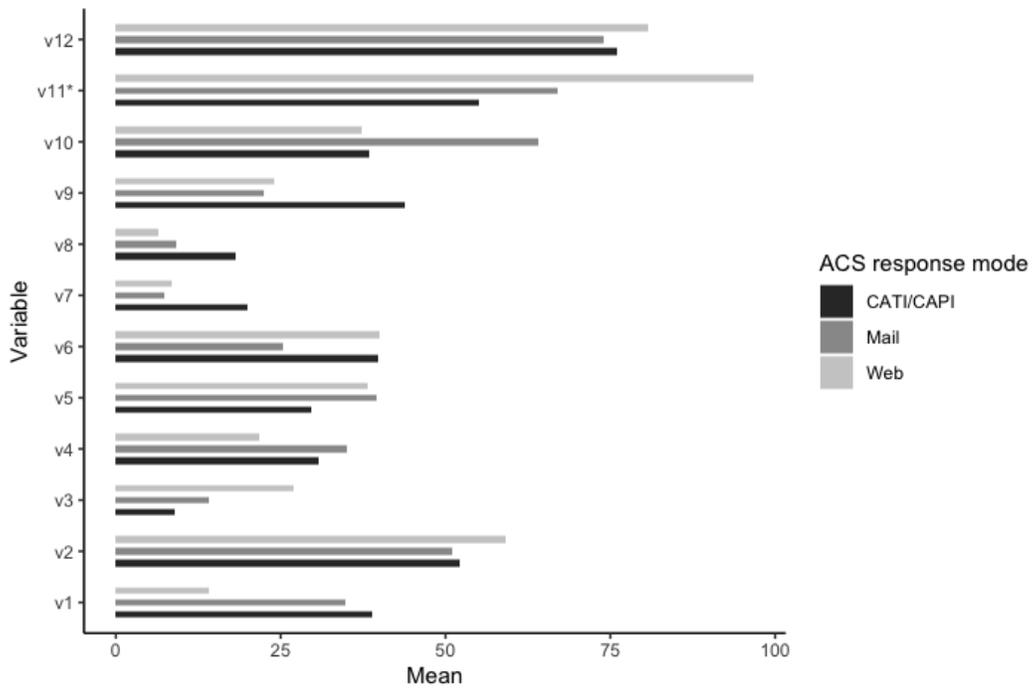

**Exhibit 4.1.  Pseudopopulation definitions**

| Pseudo-population | Pseudopopulation Response Category by ACS Response Mode | | |
|---|---|---|---|
| | **ACS Web Respondents** | **ACS Mail Respondents** | **ACS CATI/CAPI Respondents** |
| A | Web | Ftf | NR |
| B | Web | 50% Ftf, 50% NR | 50% Ftf, 50% NR |
| C | Web | Ftf | Ftf |
| D | 50% Ftf, 50% NR | Web | 50% Ftf, 50% NR |

More details on the sample sizes and subsampling fractions used in each of these designs are given in the Appendix (Exhibit A.1). These designs were constructed to yield the same expected total number of completes (3127.7) and the same number by mode.

We ran 5,000 independent iterations of each scenario. For the hybrid design scenarios, we computed each of the estimators shown in Exhibit 3.1 ($\hat{t}_1$, $\hat{t}_2$, $\hat{t}_A$, $\hat{t}_{df,1}$

, and $\hat{t}_{df,2}$). For the two subsampling scenarios, we computed $\hat{t}_1$ and $\hat{t}_2$. For the two-phase PSU subsampling scenario (B2P), we also computed the variant on the estimator $\hat{t}_2$ that uses the alternative subsampling adjustment $\omega_s^{-1}$ given in (3.6).

**Exhibit 4.2.    Simulation scenarios**

| Scenario | Pseudo-population | Design | Nonresponse follow-up |
|---|---|---|---|
| A1A | A | 1 (independent clustered and unclustered samples) | A (all) |
| B1A | B | 1 (independent clustered and unclustered samples) | A (all) |
| C1A | C | 1 (independent clustered and unclustered samples) | A (all) |
| D1A | D | 1 (independent clustered and unclustered samples) | A (all) |
| B2P | B | 2 (clustered sample only) | P (two-phase PSU subsampling) |
| B2U | B | 2 (clustered sample only) | U (two-phase unit subsampling) |

For each iteration, we computed the relative bias (RB), coefficient of variation (CV), relative root mean squared error (RRMSE), and an indicator of whether the confidence interval covered the population total (for a normal 95 percent CI) for each estimator, and averaged each of those measures across iterations. The relative bias (RB) and RRMSE were computed relative to the population parameter (i.e., by dividing the bias and RMSE, respectively, by the population total). We used Taylor series linearization to compute the variance estimates as discussed previously. The RB, CV, RRMSE, and CI coverage are the measures we used to evaluate the estimators.

## Results: Hybrid Design Scenarios

Figures 4.2 and 4.3 present the RB and RRMSE results, and Appendix Figures A.1 and A.2 show the CV and CI results, respectively, for each of the hybrid design simulation scenarios. For scenario A1A, the scenario with the most inherent bias due to the way the pseudopopulation is defined, the simulation results demonstrate that all of the estimators are biased for at least some characteristics. For characteristics such as v1, v3, v8, and v11, where bringing in the ftf follow-up is expected to reduce bias, we see that the estimators that appropriately incorporate the ftf follow-up ($\hat{t}_2$ and $\hat{t}_{df,2}$) are generally less biased; for variables such as v5, v7, and v9, where we would expect the ftf follow-up to increase bias (see Figure 1), no increase in bias is apparent for the estimators that incorporate ftf follow-up. In this scenario, CI coverage is generally very poor (well below the nominal 95 percent level) due to the bias in the estimates. The hybrid composite estimator $\hat{t}_{df,2}$ is generally comparable to the estimator $\hat{t}_2$ in terms of RRMSE.

For pseudopopulations B and D, the web respondents differ from the ftf respondents and the nonrespondents, but the assumption $\bar{Y}_N = \bar{Y}_F$ holds in expectation. A key difference between these two pseudopopulations is that the expected web response rate is considerably higher in pseudopopulation B than in pseudopopulation D (49 percent vs. 29 percent). As expected with this nonresponse model, the estimators $\hat{t}_2$ and $\hat{t}_{df,2}$ have the smallest relative bias and the best confidence interval coverage rates. For most characteristics examined, estimator $\hat{t}_{df,2}$ has the smallest relative RRMSE. The estimators that adjust the weights of web respondents for nonresponse, $\hat{t}_4$, $\hat{t}_{df,1}$, and $\hat{t}_1$, exhibit more bias and poor confidence interval coverage.

**Figure 4.2.** Relative biases of each estimator for each of the hybrid design scenarios

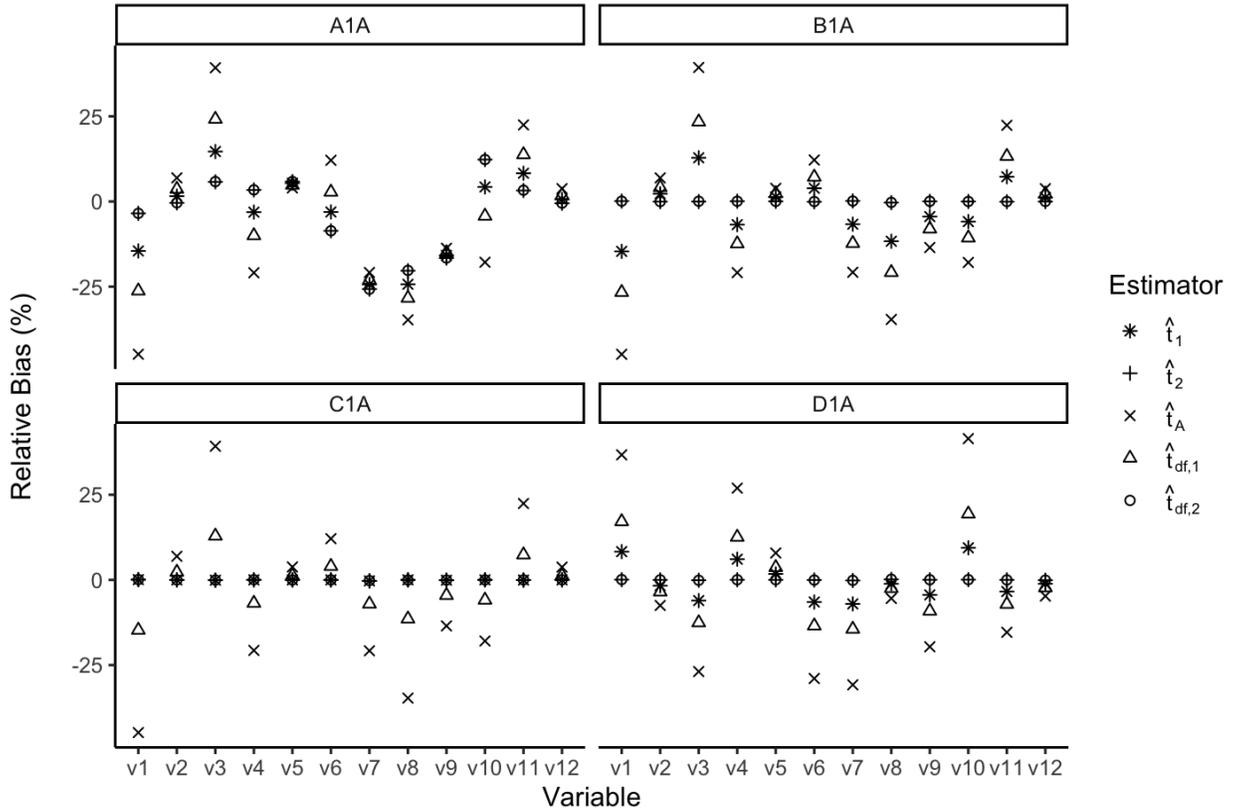

With pseudopopulation C, complete response is attained after the ftf follow-up. However, the characteristics of the web respondents generally differ from the characteristics of the ftf respondents. In scenario C1A, the relative bias is negligible for $\hat{t}_2$ and $\hat{t}_{df,2}$, the estimators that assume $\bar{Y}_N = \bar{Y}_F$. In this scenario, because $R = 1$, the estimator $\hat{t}_1$ reduces to the Horvitz-Thompson (base-weighted) estimator and exhibits properties similar to $\hat{t}_2$ and $\hat{t}_{df,2}$. Additionally, the confidence interval coverage is closest to nominal levels for these three estimators, whereas the confidence interval coverage of the other estimators is generally very poor.

**Figure 4.3.** Relative root mean squared error of each estimator for each of the hybrid design scenarios

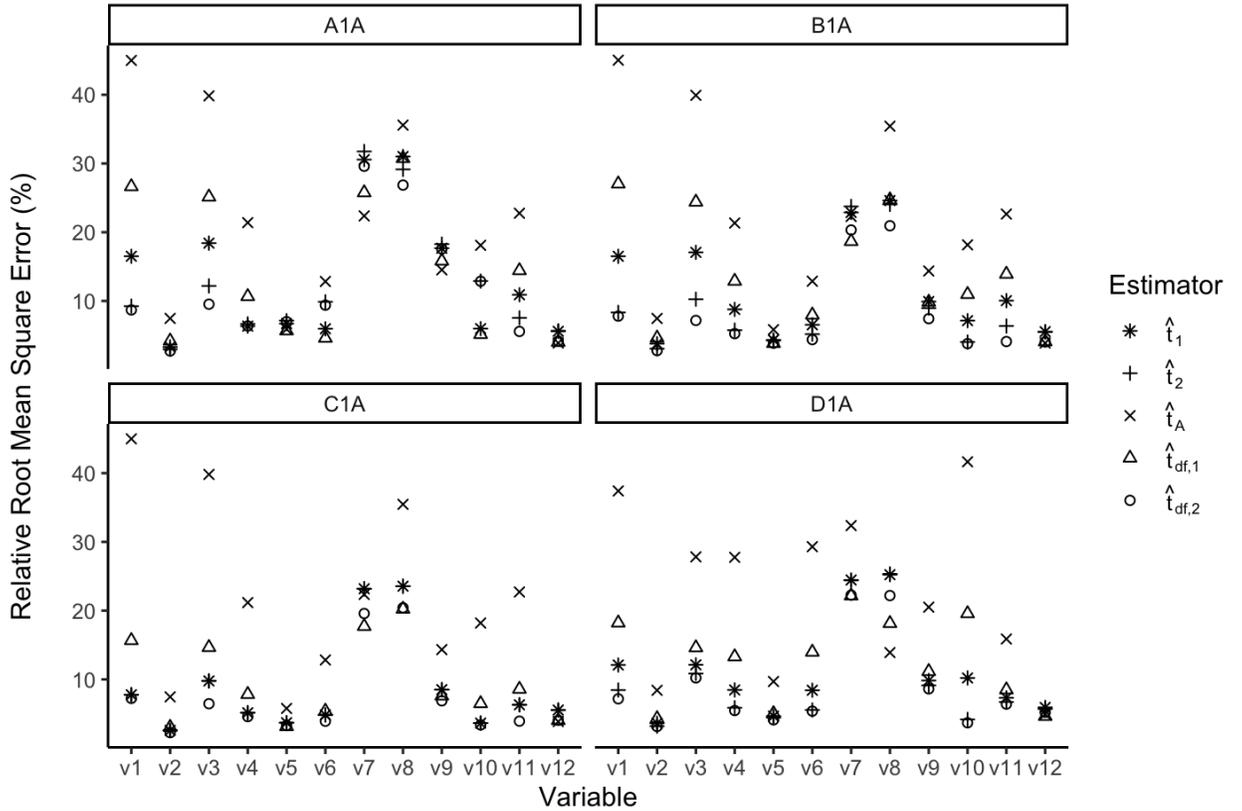

As noted in Section 1, a further consideration is the choice of compositing factors for the hybrid estimators. To examine this, we computed $\hat{t}_{df,2}$, for each iteration of the hybrid design scenarios two ways—once using the near optimal compositing factor described in Section 2 and once using the compositing factor fixed at $\kappa = 0.2$, which is far from the optimum. The results, presented in Appendix Figures A.5 – A.8, demonstrate that the choice of compositing factor has virtually no effect on the bias; for all scenarios except A1A, the same is true for the effect of the compositing factor on confidence interval coverage. The choice of compositing factor does have a small effect on CV and RRMSE, with the optimal factor resulting in less variable estimates.

## Results: Comparison of Subsampling Designs to Hybrid Design Scenario

Next, we consider the scenarios that involve subsampling for nonresponse follow-up, scenario B2U (subsampling units) and scenario B2P (subsampling PSUs). Because these scenarios involve clustered sample only, the only estimators that are relevant to these scenarios are $\hat{t}_1$ and $\hat{t}_2$. The hybrid design is an alternative to these subsampling designs, and in the previous section we demonstrated that for the hybrid design, as expected based on dual-frame estimation theory, the composite estimator $\hat{t}_{df,2}$ performed best. For all three scenarios, the estimator $\hat{t}_2$ has negligible bias while the bias in $\hat{t}_1$ is evident. (See Table A.1 in the Appendix.) Thus, our evaluation involves a comparison of the results involving $\hat{t}_2$ for scenarios B2U (subsampling units) and B2P (subsampling PSUs) to the results for $\hat{t}_{df,2}$ in scenario

B1A (hybrid design, no subsampling). For these comparisons, Figures 4.4 and 4.5 present the RB and RRMSE, and Figures A.3 and A.4 (in the Appendix) show the CV and CI coverage results.

**Figure 4.4.** Relative biases of each estimator for the hybrid design scenario vs. the subsampling scenarios

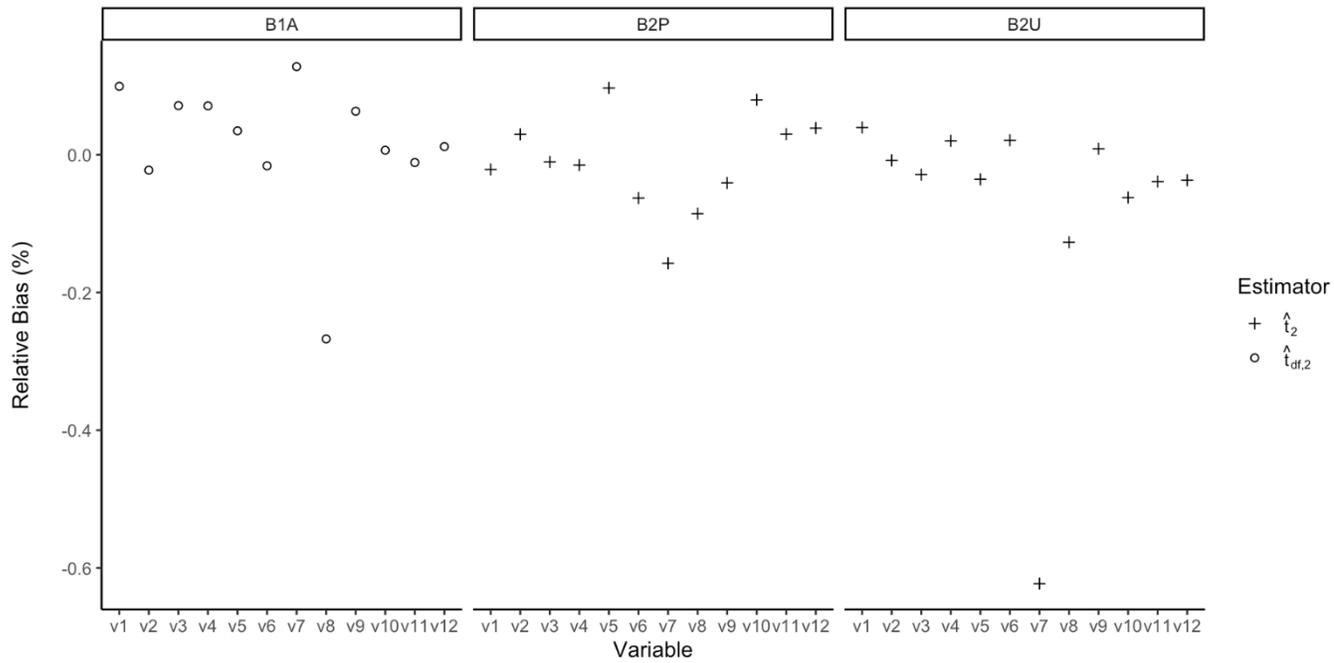

**Figure 4.5.** Relative root mean squared error of each for the hybrid design scenario vs. the subsampling scenarios

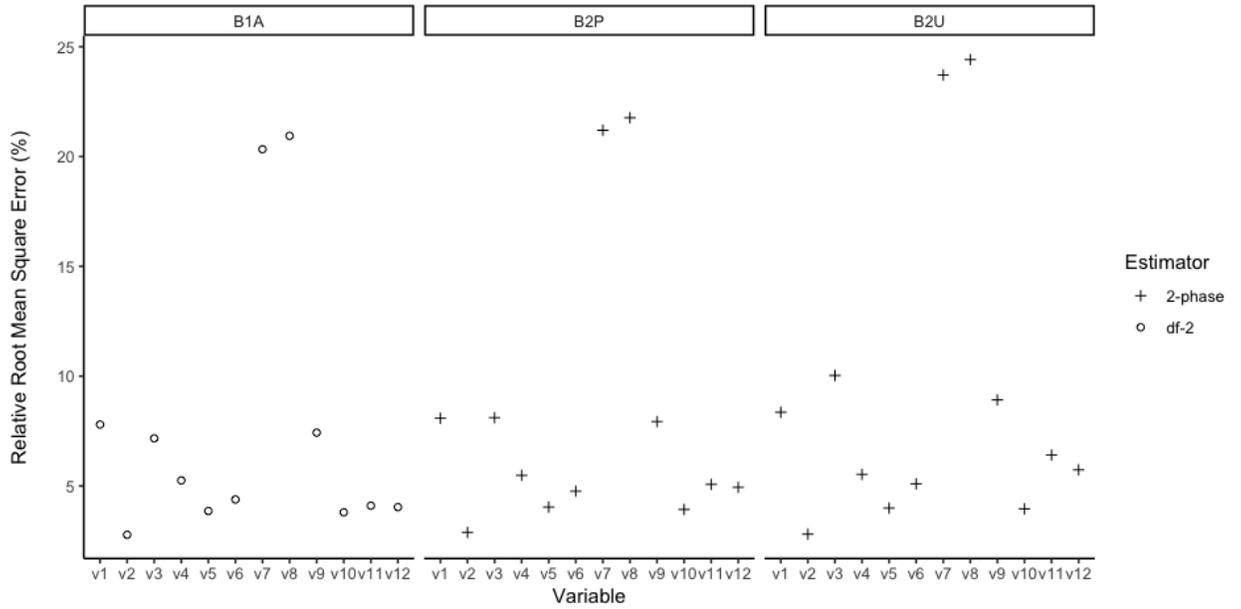

There are no appreciable differences in bias for $\hat{t}_{df,2}$ and $\hat{t}_2$ across the designs. The estimator $\hat{t}_2$ with the subsampling designs is generally more variable than the composite estimator $\hat{t}_{df,2}$ with the hybrid design, based on the CV and RRMSE. Confidence interval coverages are comparable for all three scenarios with their associated estimators.

For the PSU subsampling scenario, we evaluated the variation of the estimator $\hat{t}_2$ that uses the alternative subsampling adjustment $\omega_s^{-1}$ given in equation (3.4) by comparing it to the estimator that uses the reciprocal of the PSU subsampling rate as the adjustment $\omega^{-1}$. The results of this comparison, shown in Figures A.9 - A.12 in the Appendix, are that the estimators are comparable with respect to bias, but the alternative estimator performs marginally better with respect to CV, RRMSE, and CI coverage.

Table A.1 in the Appendix presents summaries, averaged across variables, for all of the estimators considered under each scenario.

## Section 5. Discussion

As response rates have declined and survey costs have increased, survey researchers have sought approaches to combat these trends. In this paper, we have presented design and estimation approaches for multimode data collection and have used a simulation study to examine their effectiveness. While our focus is on web for the first phase of data collection, our findings and recommendations also pertain to designs in which mail (used alone or in combination with web, as in a web-push approach) is used in the first phase.

Data collection protocols that incorporate ftf interviewing have long been held as a gold standard, typically achieving higher response rates than other modes and reducing bias by eliciting response from subgroups that are generally missed by other modes (e.g., non-telephone households missed by telephone surveys or households without internet access missed by web surveys). However, ftf interviewing is expensive relative to other modes, and as a result has been cost-prohibitive for many studies that have turned, instead, to web, phone, or paper survey administrations, or combinations of these lower-cost modes.

We presented two sample designs and associated estimators that are alternatives to two-phase unit subsampling and may open the door to ftf data collection for some studies; for other studies, the approaches we described may facilitate more discriminating use of ftf interviewing as a way of constraining costs while maintaining high standards for quality. The PSU subsampling approach uses only a clustered sample, selecting initially a larger number of PSUs, but only uses ftf follow-up in a subsample of PSUs. The estimator $\hat{t}_2$ performed well for both subsampling designs, whereas $\hat{t}_1$ exhibited substantial bias and poor CI coverage. Clearly, trying to reduce the variance by using $\hat{t}_1$ is likely to result in biases in many situations.

The hybrid design approach blends two independent samples, an unclustered sample and a clustered sample, with the less costly initial mode(s) of data collection applied to both samples but the more costly ftf nonresponse follow-up used only in the clustered sample. As a result, the hybrid design approach offers the advantage of the same number of completes at a lower cost, while achieving the same overall weighted response rate and bias reduction as a design with complete nonresponse follow-up. With the hybrid design, the hybrid estimator $\hat{t}_{df,2}$ is unbiased under the nonresponse model $\bar{Y}_N = \bar{Y}_F$ or if the ftf achieves 100% response.

With the combination of parameters used in the simulation study, the composite estimator $\hat{t}_{df,2}$ in the hybrid design scenario had a RRMSE that was, on average, 8 percent lower than $\hat{t}_2$ in the PSU subsampling design scenario and 14 percent lower than $\hat{t}_2$ in the unit subsampling design scenario. All three scenarios were designed to have about the same cost by having the same expected number of completes by mode and with ftf follow-up administered in the same number of PSUs.

The simulation findings suggest that the hybrid design and $\hat{t}_{df,2}$ is superior to either of the two-phase designs in the conditions we examined, and the two-phase PSU subsampling is superior to the two-phase unit subsampling. In addition, the hybrid design has practical benefits, especially when there is uncertainty regarding response rates by mode. With the hybrid design, increasing the sample size in the unclustered sample to deal with a lower-than-expected web response rate is relatively simple. With the two-phase designs, there are more complications such as the cost and schedule implications of the ftf follow-up due to increasing the first-phase sample.

The simulation consistently demonstrated the well-known shortcomings of the estimator $\hat{t}_1$ that is commonly used in practice, where there is no distinction in respondents by mode during the computation of weighting adjustments. Implicit in this estimator is the nonresponse model that assumes equality of means among the web respondents, the ftf respondents, and the nonrespondents. This is more restrictive than the model of equality of means between the ftf respondents and the nonrespondents that underlies the estimator $\hat{t}_2$ and the hybrid estimator $\hat{t}_{df,2}$. Brick et al. (2021) discuss this issue and partially account for imbalances in the respondent composition by using an adaptive mode adjustment. In a study involving web with ftf follow-up, the Brick et al. (2021) approach may be implemented by adjusting both the web and ftf respondents' weights to retain some of the bias reduction qualities of the adjustment of the ftf respondents while reducing variances by adjusting the weights of the web respondents as well. More research is needed to explore this approach with ftf surveys.

We have included discussion of theoretical properties of the estimators in the context of the sample designs presented here. However, as with any simulation, we have not examined every possible scenario. In designing our simulation, we focused on aspects we believed to be most likely to affect the relative performance of the estimators.

The estimators we presented incorporate a single adjustment, effectively treating the sample as a single weighting class. In practice, we would not expect the assumption $\bar{Y}_N = \bar{Y}_F$ to hold in general, but this assumption might be better approximated within classes, i.e., $\bar{Y}_{N_C} = \bar{Y}_{F_C}$, where the subscript *C* denotes the class. If auxiliary variables can be identified and are available for both respondents and nonrespondents such that this nonresponse model holds (at least approximately), then these classes would typically be used in computing the weighting adjustments in order to reduce bias. Further research is needed to extend the estimators we presented to this situation, using weighting class-specific adjustments for nonresponse, for both the hybrid design and the PSU subsampling design. Similarly, calibrated estimators need to be evaluated.

We conclude with a few thoughts about another design we believe warrants further examination—an unclustered design in combination with a sample of existing PSUs – where the design that begins with web and uses ftf follow-up of the web nonrespondents in the sampled PSUs. One possible application of this design is for sample replenishment in a longitudinal study in which the original sample was a clustered sample of PSUs. Another application is where trained staff are available only in an existing sample of PSUs. In such contexts, one could consider selecting an unclustered sample, attempting the survey by web first, and using ftf follow-up for the web nonrespondents in the unclustered sample that are located within the particular PSUs. We believe the extension of the estimator $\hat{t}_2$ shown in (3.5), with a modification to the second term so that the $\omega^{-1}$ is replaced by the reciprocal of the PSU probability of selection, is suitable in this context. Further work is needed to more fully examine this design and the properties of the estimator in this context.

## Acknowledgements

We are grateful to Sharon Lohr for her very helpful thoughts and comments on a draft of this manuscript. We also greatly appreciate the comments and suggestions offered by the Editor, Associate Editor, and three reviewers.

# Appendix

**Exhibit A.1. Parameters for nonresponse follow-up designs in simulation**

|  | Total for unclustered sample | Total for clustered sample | |
|---|---|---|---|
|  |  | Per PSU | Overall |
| **No subsampling for NRFU (Scenario B1A)** | | | |
| Unclustered sample size | 2500 | | |
| # PSUs | | | 50 |
| # units sampled per PSU | | | 50 |
| Expected # completes | | | |
|   Web | | 24.0 | 1201.0 |
|   Ftf (in all 50 PSUs) | | 14.5 | 725.7 |
|   Total | 1201.0 | 38.5 | 1926.7 |
| **Two-phase unit subsampling (Scenario B2U)** | | | |
| Unclustered sample size | 0 | | |
| # PSUs | | | 50 |
| # units sampled per PSU | | | 100 |
|   Nonresponse follow-up subsampling fraction | | | 0.5 |
| Expected # completes | | | |
|   Web | | 48.0 | 2402.1 |
|   Ftf (in all 50 PSUs) | | 14.5 | 725.7 |
|   Total | 0 | 62.6 | 3127.7 |
| **Two-phase PSU subsampling (Scenario B2P)** | | | |
| Unclustered sample size | 0 | | |
| # PSUs | | | |
|   total PSUs in sample | | | 100 |
|   PSUs subsampled for NRFU | | | 50 |
| # units sampled per PSU | | | 50 |
| Expected # completes per PSU | | | |
|   Web | | 24.0 | 2402.1 |
|   Ftf (in only the 50 PSUs subsampled for NRFU) | | 14.5 | 725.7 |
|   Total in subsampled PSUs | | 38.5 | |
| Expected total # completes | 0 | | 3127.7 |

**Table A.1.    Mean of summary measures of estimators by scenario**

| Statistic | Scenario | 2-phase | Unclustered | 2-stage | df-1 | df-2 (opt) | df-2 (not opt) | 2-phase (PS) |
|---|---|---|---|---|---|---|---|---|
| RB | A1A | -3.76% | -5.36% | -4.20% | -4.66% | -3.73% | -3.76% | |
|  | B1A | -0.02% | -5.34% | -1.78% | -3.19% | 0.01% | -0.01% | |
|  | C1A | -0.03% | -5.35% | -0.03% | -1.79% | -0.01% | -0.03% | |
|  | D1A | 0.00% | -2.19% | -0.49% | -1.02% | 0.01% | 0.00% | |
|  | B2P | -0.01% |  | -1.78% |  |  |  | -0.02% |
|  | B2U | -0.07% |  | -1.82% |  |  |  | |
| CV | A1A | 7.31% | 4.28% | 7.19% | 4.74% | 5.95% | 6.69% | |
|  | B1A | 8.16% | 4.27% | 7.70% | 4.96% | 6.82% | 7.55% | |
|  | C1A | 7.79% | 4.27% | 7.79% | 5.44% | 6.39% | 7.15% | |
|  | D1A | 8.51% | 5.61% | 8.23% | 5.98% | 7.76% | 8.17% | |
|  | B2P | 7.37% |  | 6.39% |  |  |  | 7.18% |
|  | B2U | 8.15% |  | 7.54% |  |  |  | |
| RRMSE | A1A | 12.78% | 20.81% | 13.25% | 14.40% | 11.69% | 12.27% | |
|  | B1A | 9.12% | 20.77% | 11.43% | 13.58% | 7.66% | 8.45% | |
|  | C1A | 8.71% | 20.75% | 8.71% | 9.53% | 7.16% | 8.00% | |
|  | D1A | 9.48% | 22.47% | 11.03% | 12.80% | 8.66% | 9.11% | |
|  | B2P | 8.18% |  | 10.27% |  |  |  | 7.97% |
|  | B2U | 9.08% |  | 11.33% |  |  |  | |

**Table A.1.    Mean of summary measures of estimators by scenario (Cont.)**

| Statistic | Scenario | 2-phase | Unclustered | 2-stage | df-1 | df-2 (opt) | df-2 (not opt) | 2-phase (PS) |
|---|---|---|---|---|---|---|---|---|
| CI | A1A | 72.58% | 12.86% | 75.06% | 39.85% | 65.24% | 68.93% | |
| | B1A | 94.96% | 12.89% | 81.40% | 39.46% | 94.66% | 94.60% | |
| | C1A | 94.05% | 12.95% | 94.05% | 63.57% | 93.42% | 93.43% | |
| | D1A | 94.70% | 18.48% | 83.41% | 49.66% | 94.36% | 94.30% | |
| | B2P | 94.72% | | 75.19% | | | | 95.31% |
| | B2U | 94.91% | | 79.97% | | | | |
| ABS(RB) | A1A | 8.85% | 20.12% | 9.99% | 13.16% | 8.86% | 8.85% | |
| | B1A | 0.09% | 20.08% | 6.57% | 11.95% | 0.07% | 0.08% | |
| | C1A | 0.05% | 20.06% | 0.05% | 6.63% | 0.05% | 0.05% | |
| | D1A | 0.06% | 21.03% | 4.75% | 9.81% | 0.08% | 0.06% | |
| | B2P | 0.06% | | 6.64% | | | | 0.07% |
| | B2U | 0.09% | | 6.64% | | | | |
| NormCIL | A1A | 1.16 | 0.71 | 1.16 | 0.76 | 0.93 | 1.04 | |
| | B1A | 1.20 | 0.71 | 1.18 | 0.76 | 0.99 | 1.08 | |
| | C1A | 1.08 | 0.71 | 1.08 | 0.75 | 0.85 | 0.96 | |
| | D1A | 1.23 | 0.93 | 1.18 | 0.85 | 1.12 | 1.16 | |
| | B2P | 1.07 | | 0.95 | | | | 1.07 |
| | B2U | 1.19 | | 1.13 | | | | |

For the CIL, values for each variable divided by the mean of the CIL across the scenarios

**Figure A.1.     Coefficients of variation of each estimator for each of the hybrid design scenarios**

Variable definitions: v1-high school graduate; v2-some college; v3-Bachelor's or higher; v4-1-person household; v5-2-person household; v6-3 or more person household; v7-Hispanic reference person; v8-Black, nonHispanic reference person; v9-renter; v10-person 60 years or older; v11*-household income/110000; v12-percent urban.

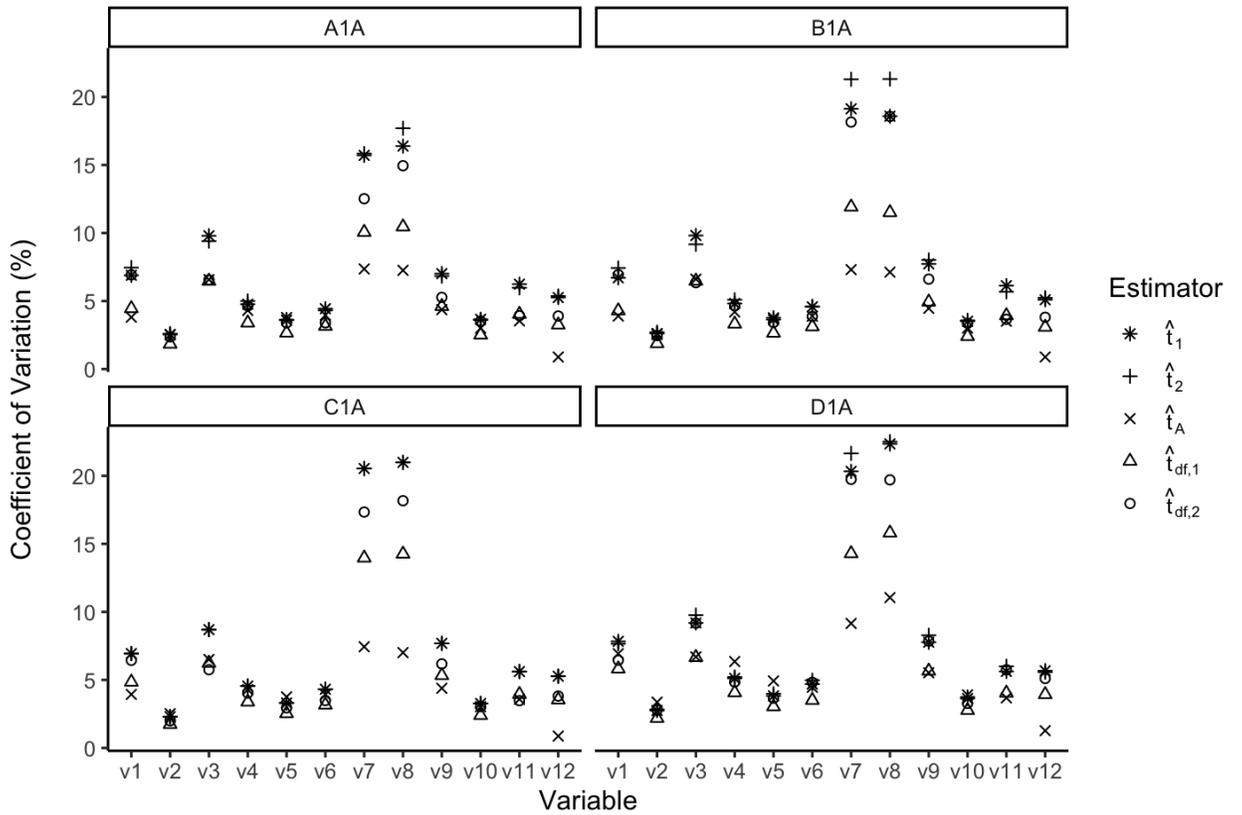

**Figure A.2.** Confidence interval coverage for each estimator for each of the hybrid design scenarios

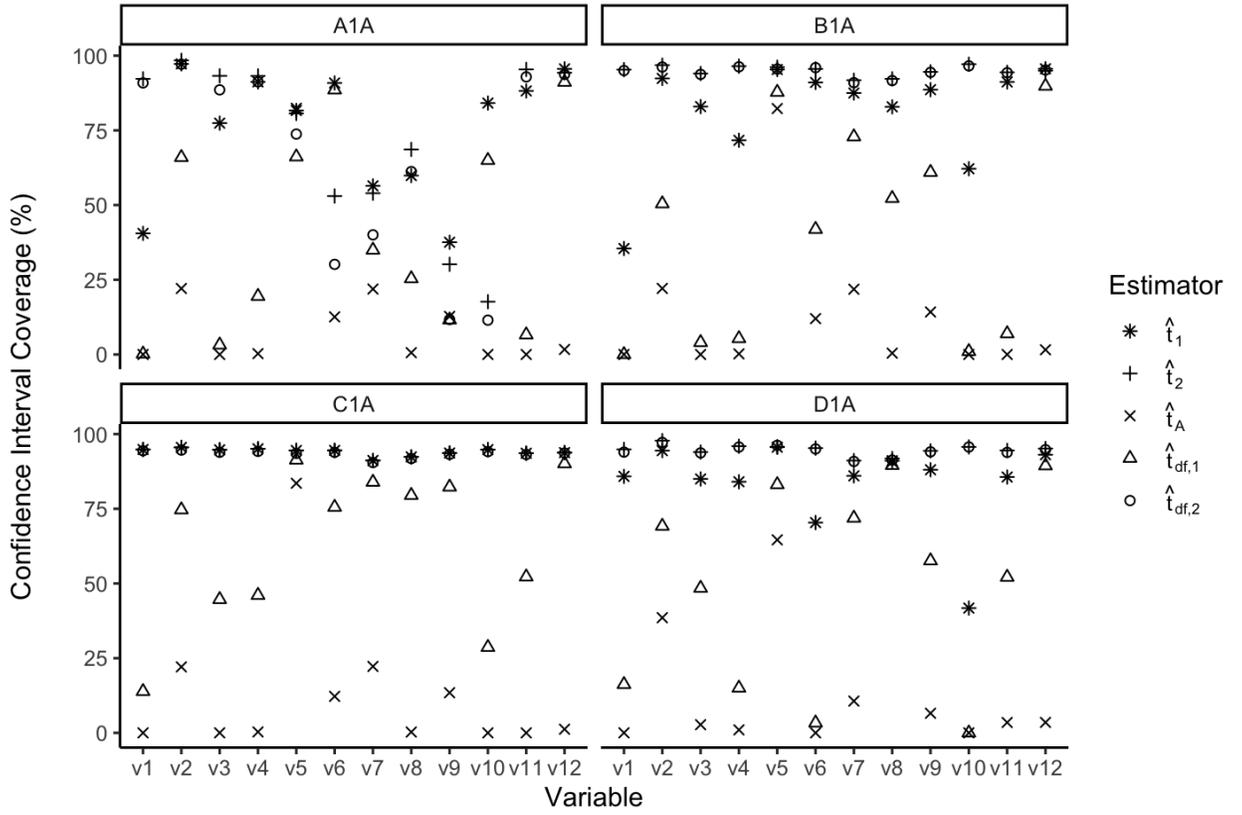

**Figure A.3.** Coefficients of variation of each estimator for the hybrid design scenario vs. the subsampling scenarios

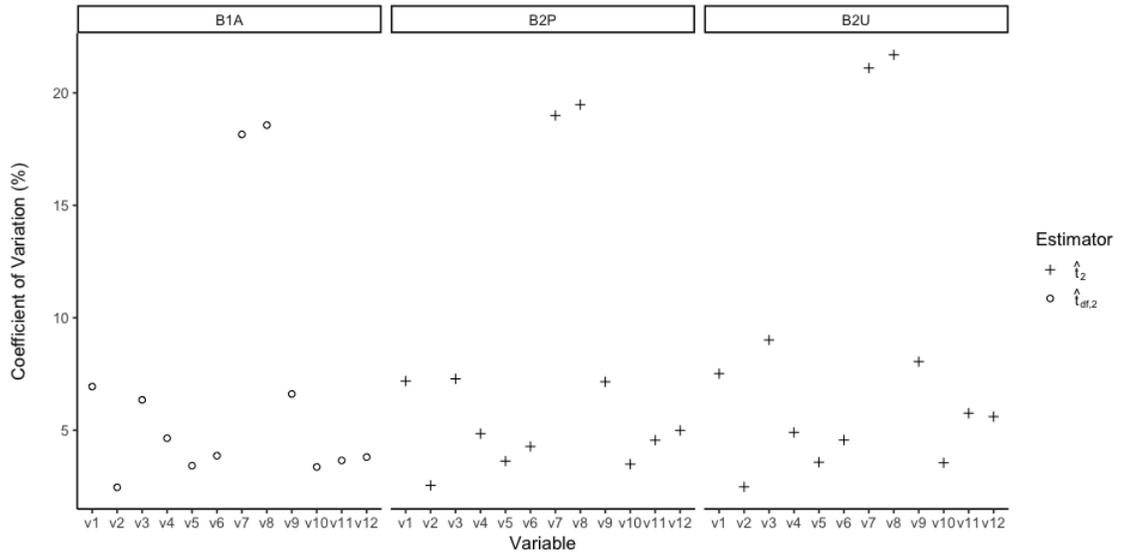

**Figure A.4.    Confidence interval coverage for each for the hybrid design scenario vs. the subsampling scenarios**

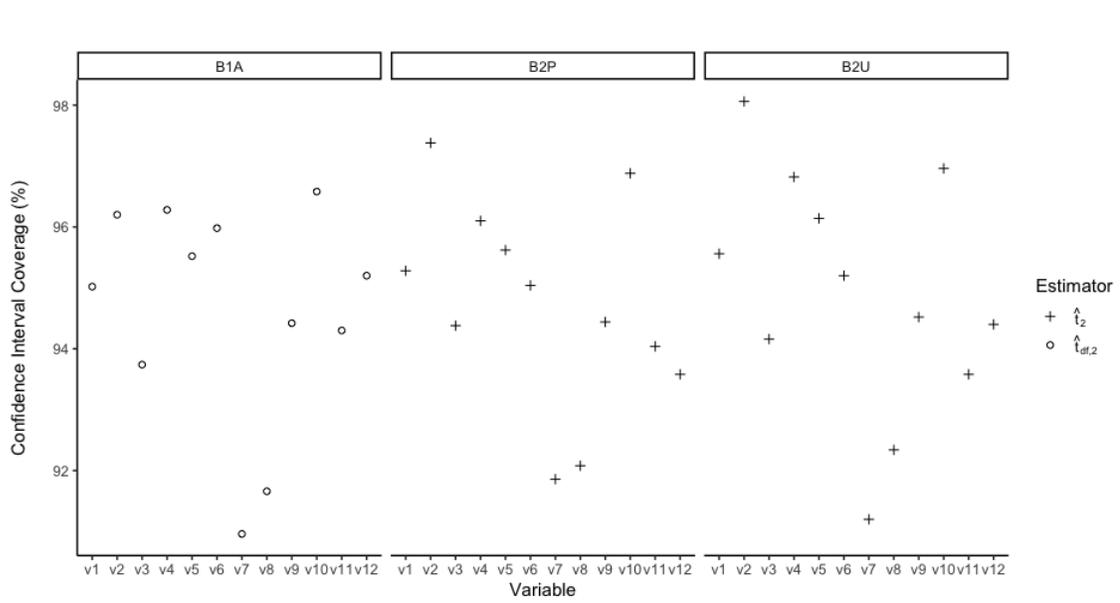

**Figure A.5.** Relative biases of the hybrid composite estimator $\hat{t}_{df,2}$ based on both optimal and non-optimal compositing factors

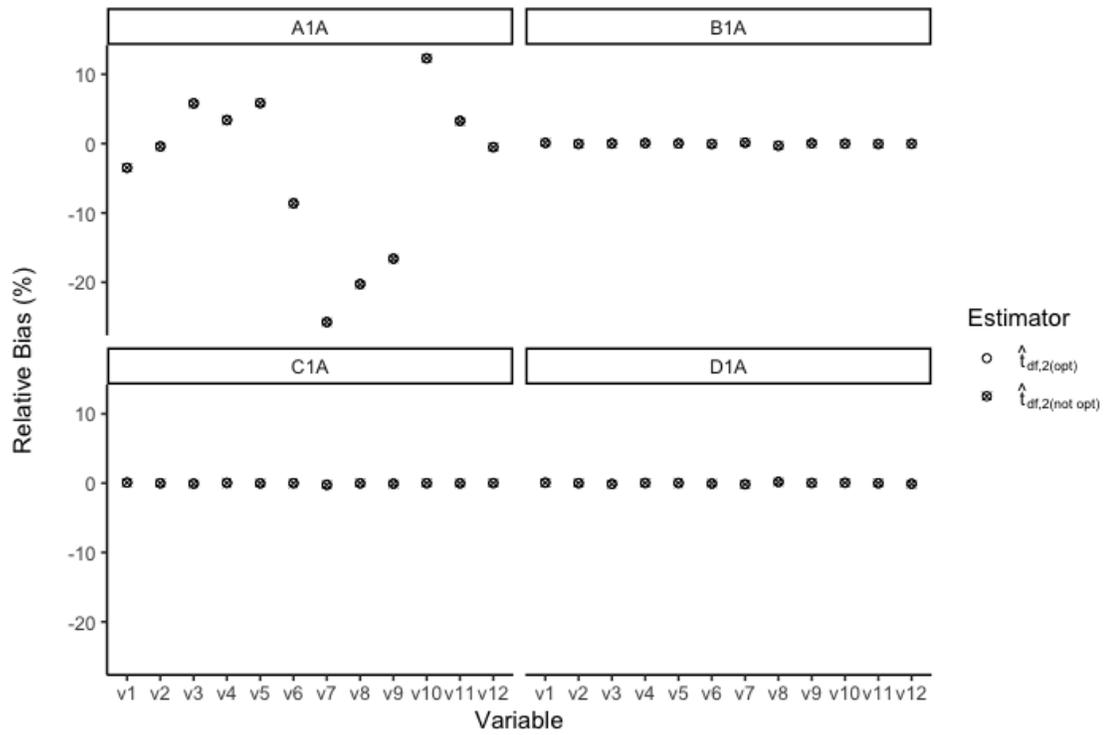

**Figure A.6.** Coefficients of variation of the hybrid composite estimator $\hat{t}_{df,2}$ based on both optimal and non-optimal compositing factors

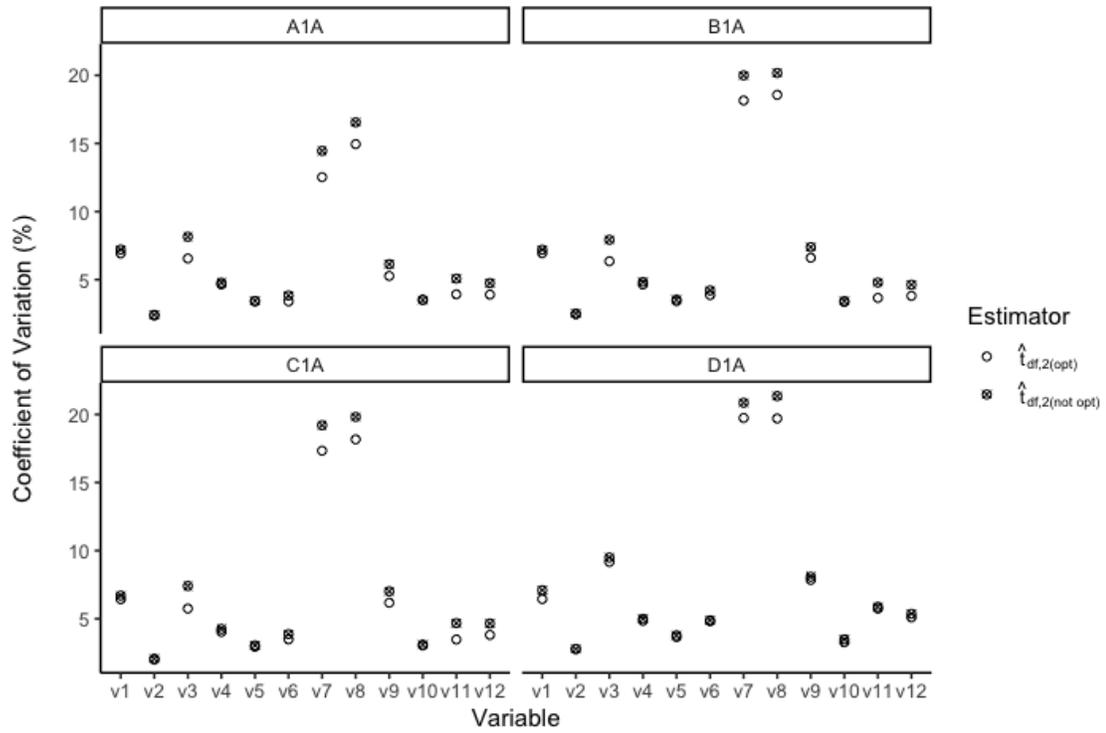

**Figure A.7.** Relative root mean squared error of the hybrid composite estimator $\hat{t}_{df,2}$ based on both optimal and non-optimal compositing factors

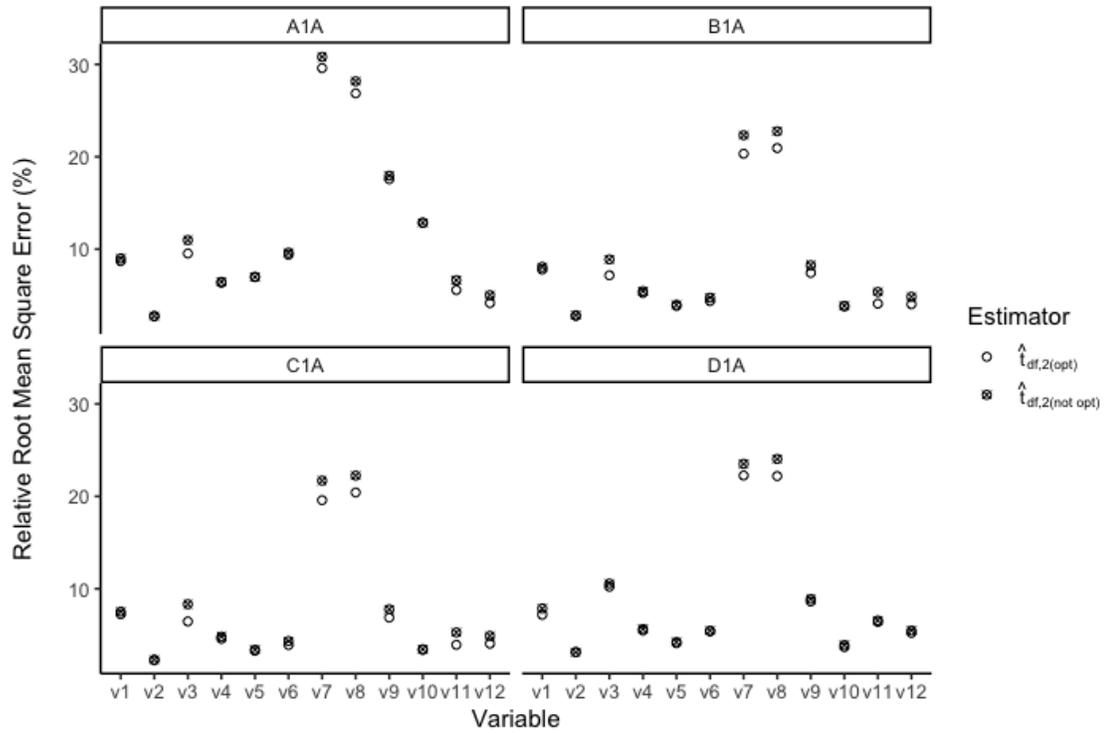

**Figure A.8.** Confidence interval coverage of the hybrid composite estimator $\hat{t}_{df,2}$ based on both optimal and non-optimal compositing factors

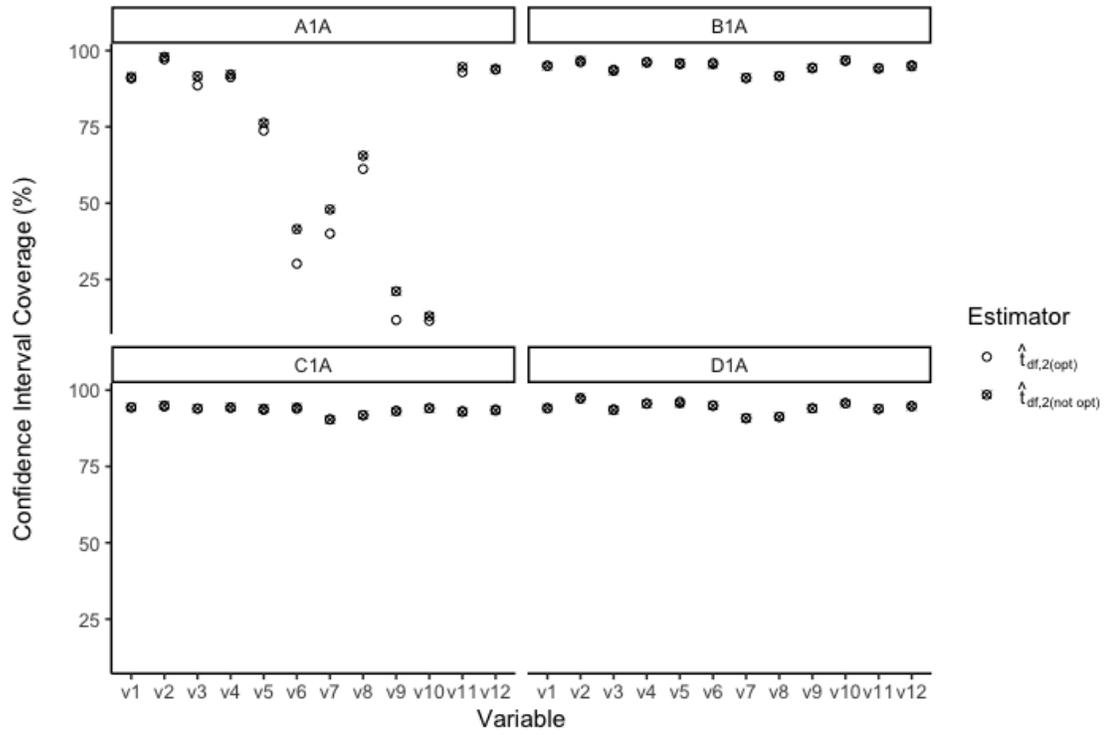

**Figure A.9.** Relative biases of two versions of the estimator $\hat{t}_2$ for PSU subsampling

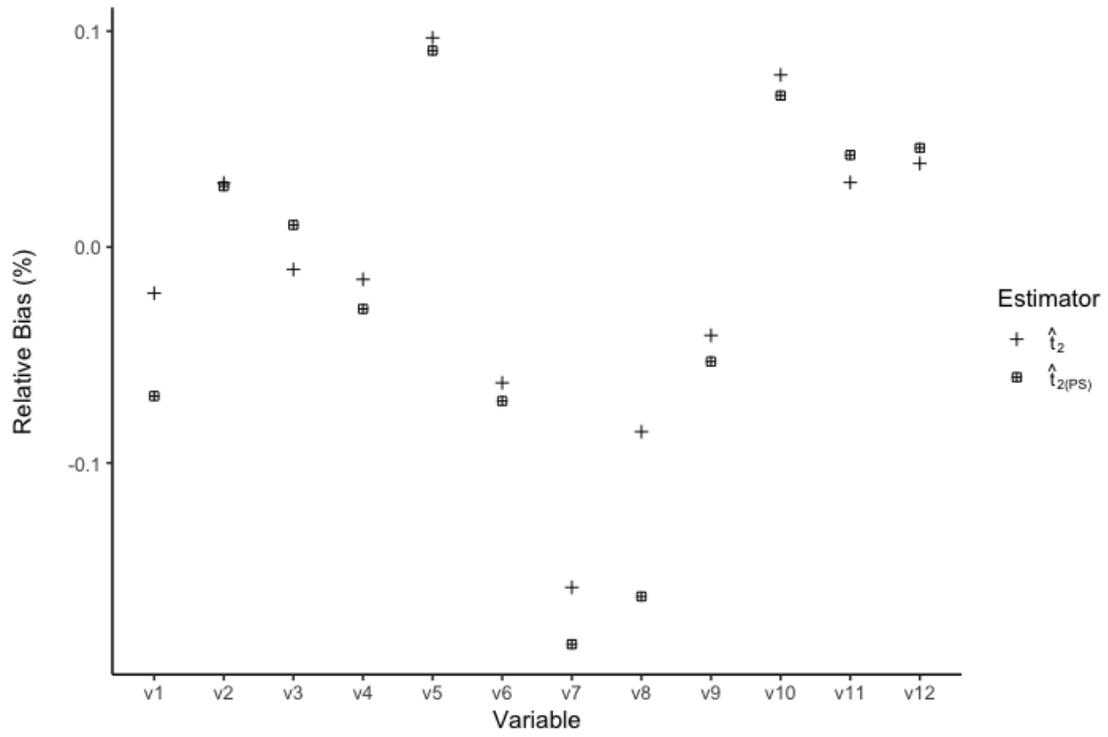

**Figure A.10. Coefficients of variation of two versions of the estimator $\hat{t}_2$ for PSU subsampling**

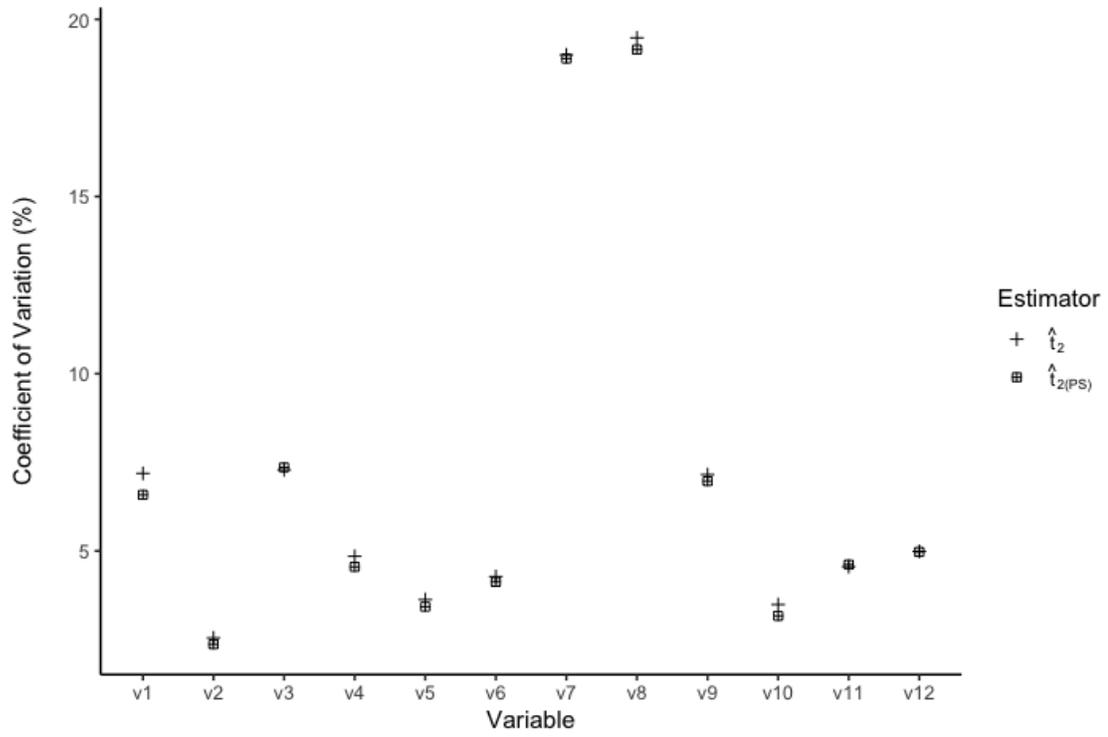

**Figure A.11.** Relative root mean squared error of two versions of the estimator $\hat{t}_2$ for PSU subsampling

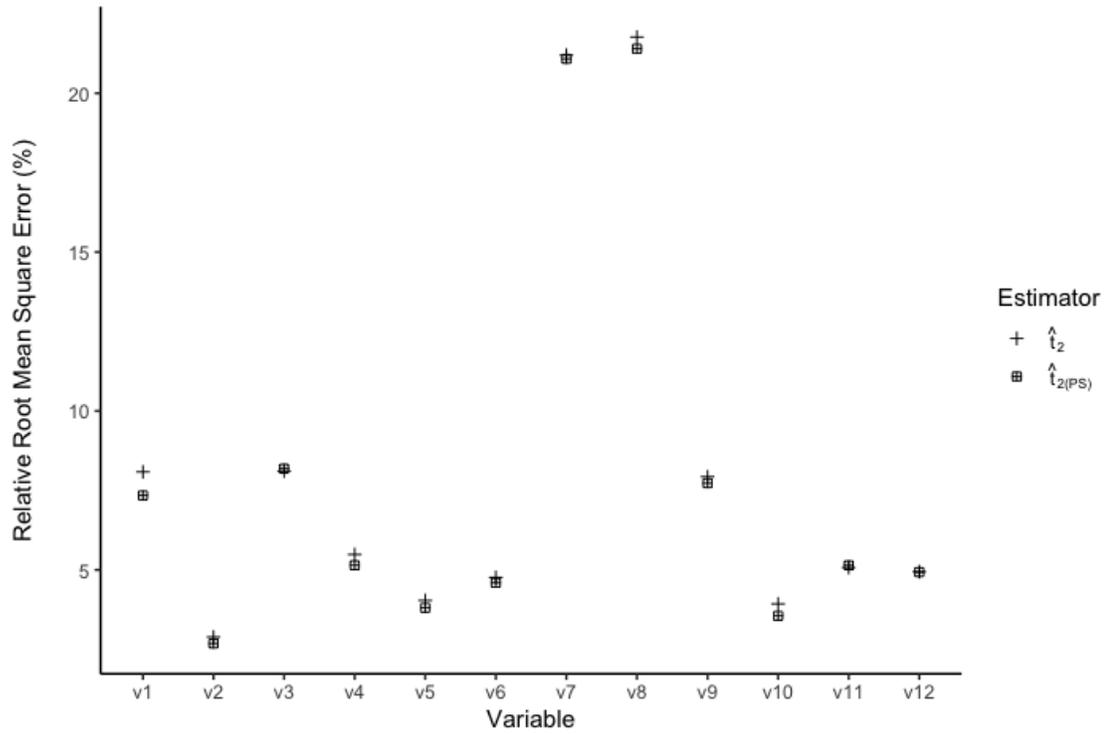

**Figure A.12.** Confidence interval coverage of two versions of the estimator $\hat{t}_2$ for PSU subsampling

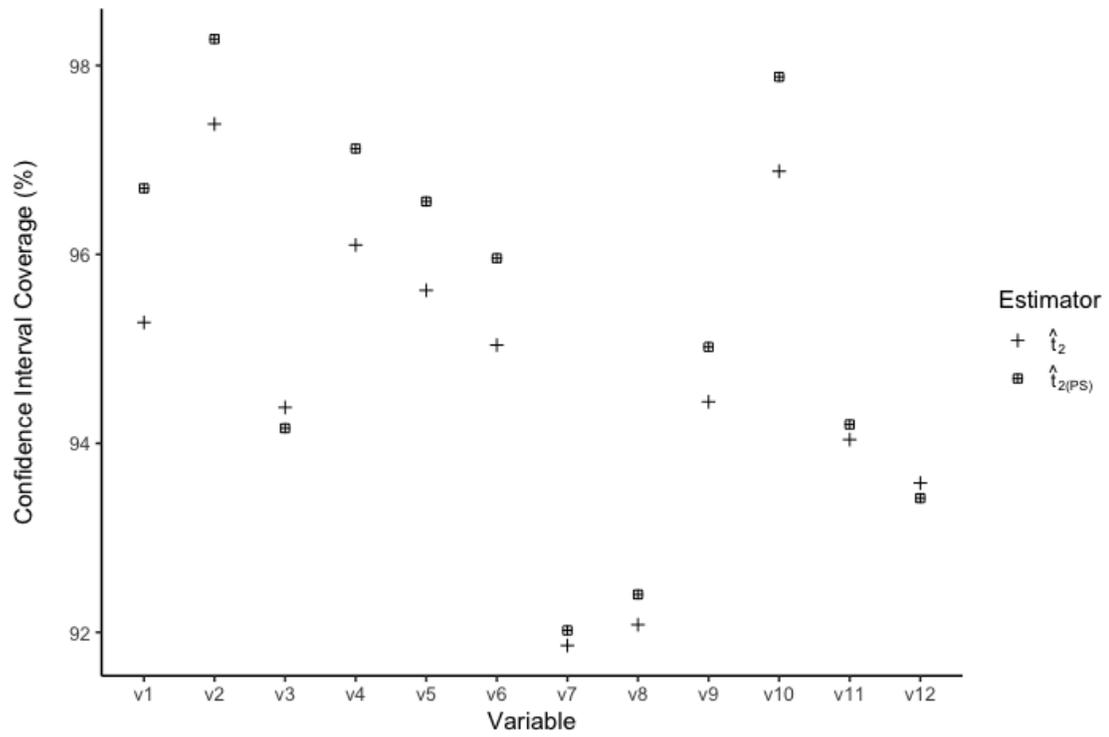